\documentclass[runningheads,anonymous]{llncs}

\usepackage{amsfonts}
\usepackage{amsmath}
\usepackage{multirow}
\usepackage{graphicx}
\usepackage{booktabs}
\usepackage{marvosym}

\begin{document}
\title{AsynFusion: Towards Asynchronous Latent Consistency Models for Decoupled Whole-Body Audio-Driven Avatars}
%
%

\author{Tianbao Zhang\inst{1}\textsuperscript{\textdagger}  \and
Jian Zhao\inst{1}\textsuperscript{\textdagger} \and
Yuer Li\inst{1} \and
Zheng Zhu\inst{5} \and
Ping Hu\inst{4} \and
Zhaoxin Fan\inst{2,3}\textsuperscript{\Letter} \and
Wenjun Wu\inst{2,3} \and
Xuelong Li\inst{1}\textsuperscript{\Letter}}
%
%
\institute{Institute of Artificial Intelligence (TeleAI), China Telecom \and
Beijing Advanced Innovation Center for Future Blockchain and Privacy Computing, School of Artificial Intelligence, Beihang University \and
Hangzhou International Innovation Institute, Beihang University \and
School of Computer Science and Technology, Xinjiang University \and
GigaAI}


\maketitle

\begingroup
\renewcommand\thefootnote{\textdagger}
\footnotetext{Equal Contribution.}
\renewcommand\thefootnote{\Letter}
\footnotetext{Corresponding authors.}
\endgroup


\begin{abstract}
Whole-body audio-driven avatar pose and expression generation is a critical task for creating lifelike digital humans and enhancing the capabilities of interactive virtual agents, with wide-ranging applications in virtual reality, digital entertainment, and remote communication. Existing approaches often generate audio-driven facial expressions and gestures independently, which introduces a significant limitation: the lack of seamless coordination between facial and gestural elements, resulting in less natural and cohesive animations. To address this limitation, we propose AsynFusion, a novel framework that leverages diffusion transformers to achieve harmonious expression and gesture synthesis. The proposed method is built upon a dual-branch DiT architecture, which enables the parallel generation of facial expressions and gestures. Within the model, we introduce a Cooperative Synchronization Module to facilitate bidirectional feature interaction between the two modalities, and an Asynchronous LCM Sampling strategy to reduce computational overhead while maintaining high-quality outputs. Extensive experiments demonstrate that AsynFusion achieves state-of-the-art performance in generating real-time, synchronized whole-body animations, consistently outperforming existing methods in both quantitative and qualitative evaluations.

\end{abstract}

\section{Introduction}
Audio-driven avatar expression and pose generation \cite{23,2,5} is a crucial task aimed at creating lifelike digital humans that can seamlessly translate audio input into synchronized facial expressions and body poses. This task is fundamental to bridging the gap between speech and nonverbal communication, enabling avatars to convey emotions, intentions, and personality in a natural and dynamic manner. Its importance spans a wide range of fields, including metaverse applications, digital human development, gaming, and human-computer interaction \cite{48,meta,meta2}.  

\begin{figure}
\vspace{-0.5cm}
\includegraphics[height=0.75\textwidth,width=\textwidth]{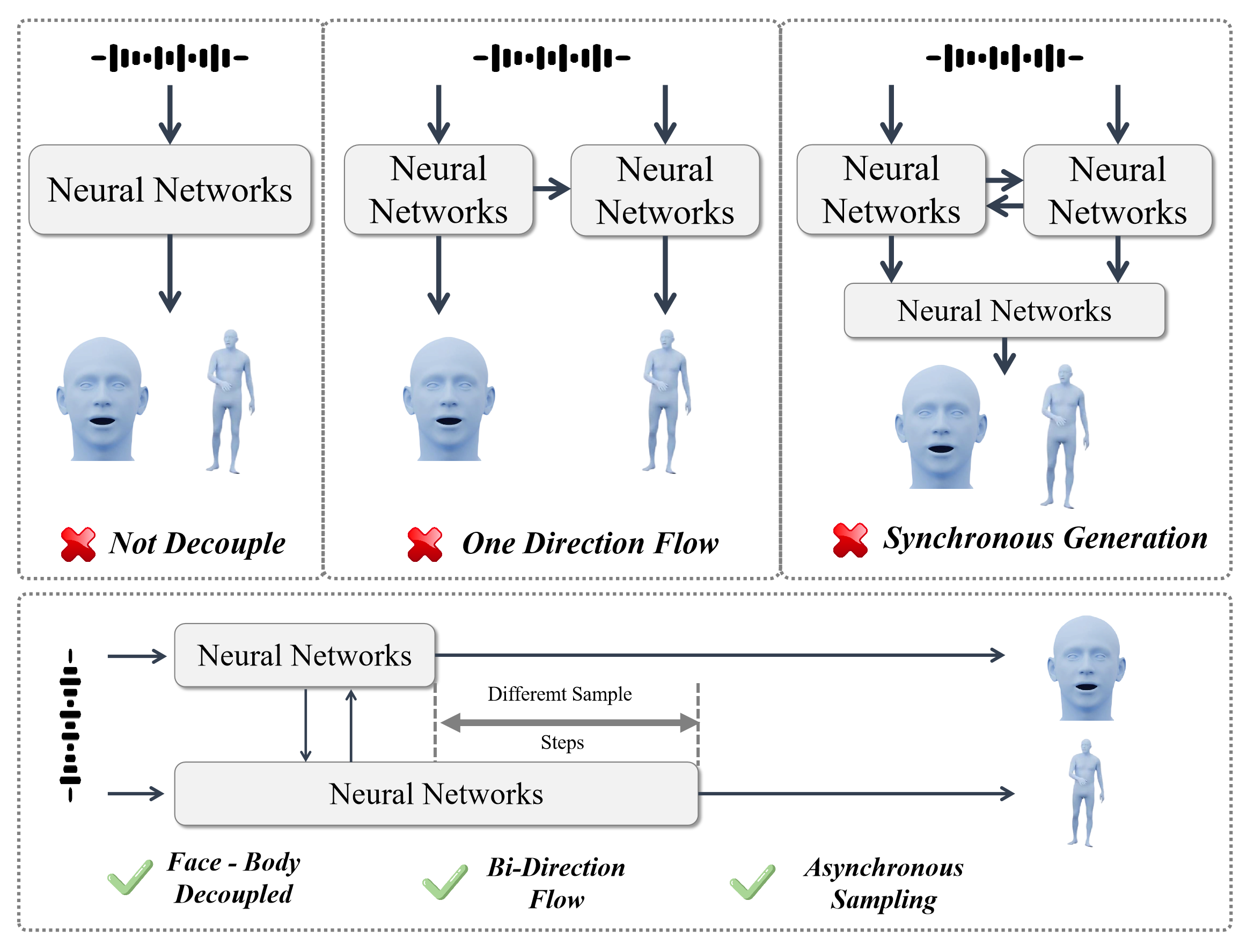}
\caption{Comparison of Different Audio-Driven Avatar Generation Frameworks. The upper section presents the three mainstream frameworks, while the lower section introduces our proposed AsynFusion which enables bidirectional feature interaction between the face and body generators and supports asynchronous sampling for more efficient generation.} \label{fig1}
\vspace{-0.5cm}
\end{figure}

In recent years, numerous methods have been proposed for audio-driven avatar expression and pose generation, primarily treating speech-driven facial expression and body motion synthesis as separate tasks. Facial expression generation \cite{face1,face2} focuses on mapping emotional features from speech to facial muscle movements for natural animations, while body motion synthesis \cite{body1,body2} explores correlations between speech and gestures to generate coherent full-body motions. Despite advancements, these methods often lack sufficient coordination between expressions and movements. Generative models like VQ-VAE \cite{2}, GANs \cite{14}, and diffusion models \cite{1,2,47,50} have improved synchronization and diversity, enabling unified modeling of expressions and movements \cite{14,31,probtalk,EMAGE}. As shown in Fig. \ref{fig1}, recent works include Probtalk \cite{probtalk}, which generates expressions and postures simultaneously with a unified model, DiffSHEG \cite{31}, which uses a unidirectional sequence from expression to gestures, and EMAGE \cite{EMAGE}, which incorporates body hints for better coordination. Combo \cite{combo}, the most related work, combines features for expressions and movements into a joint bidirectional distribution. However, a key challenge remains: balancing coordination accuracy and computational efficiency. More specifically, synchronization of expressions and movements often incurs high computational overhead, limiting the production of fluid animations in latency-sensitive scenarios. 

To address this challenge, we propose AsynFusion, a framework that decouples facial expression and body gesture generation for efficient, lifelike animation. By separating head and body generation, AsynFusion enables parallel processing while maintaining coordination through shared feature interactions. This design respects the distinct dynamics of each modality and incorporates asynchronous mechanisms to improve efficiency without sacrificing quality. The model comprises three key components: (1) a dual-branch Diffusion Transformer for parallel expression-gesture generation with bidirectional interaction; (2) a cooperative synchronization module using cross-attention to capture inter-modal dependencies and enhance coherence; and (3) an asynchronous Latent Consistency Model (LCM) sampling strategy that accelerates inference while preserving motion quality, enabling real-time applications.

Extensive experiments on widely-used benchmarks demonstrate that AsynFusion can achieve state-of-the-art performance on both generation quality and computational efficiency. Our main contributions are as follows:
\begin{itemize}
    \item We propose a novel dual-branch DiT architecture that incorporates the concept of asynchronous diffusion, enabling the parallel generation of facial expressions and body gestures. 
    
    \item We propose a cooperative synchronization module and an Asynchronous LCM-based sampling strategy to efficiently model the complex dependencies between facial expressions and body gestures while reducing computational overhead. 
    
    \item We conduct extensive quantitative and qualitative experiments to demonstrate the effectiveness, efficiency, and real-time capability of AsynFusion in generating coherent and lifelike animations.
\end{itemize}

\section{Related Work}
\subsection{Speech-driven Expression Generation}
Speech-driven facial animation has evolved from early rule-based methods \cite{22,23}, which offered controllability but required manual tuning, to data-driven models \cite{21,28,29} that generate more natural and speech-synchronized expressions. However, these models often suffer from pixel-level artifacts and geometric inconsistencies. Recent advances in 3D facial animation, especially transformer-based architectures like FaceFormer \cite{12} and CodeTalker \cite{41}, have improved temporal alignment using attention mechanisms. Still, most approaches rely on deterministic mappings, limiting expression diversity. Our work builds on these developments by introducing a more expressive framework that overcomes the limitations of deterministic designs.

\subsection{Speech-driven Gesture Generation}
Gesture generation has similarly transitioned from rule-based systems \cite{18,23} to data-driven methods using MLPs, CNNs, RNNs \cite{14,15,27}, and Transformers \cite{6}. Recognizing the one-to-many nature of speech-to-gesture mapping, recent works have adopted generative models like GANs \cite{14} and diffusion models \cite{1,31}, which offer greater motion diversity. Pioneering diffusion-based approaches such as DiffGesture \cite{DiffGesture} and DiffuseStyleGesture \cite{Diffusestylegesture} require motion seeds, limiting their use in continuous generation. LDA \cite{1} addressed longer sequences with translation-invariant embeddings but struggles with streaming data. Our method advances this line by enabling seed-free, continuous gesture generation in a more robust and efficient framework.

\subsection{Joint Expression-Gesture Generation}
Joint modeling of expressions and gestures aims to improve realism and synchronization. Habibie et al. \cite{14} proposed a CNN-based multi-decoder system with adversarial training, though it lacked motion diversity. Yi et al. \cite{show} used Wav2Vec \cite{4} and VQ-VAE to decouple tasks, but tokenization constrained gesture variety. ProboTalk \cite{probtalk} unified expression and pose generation in one model, while EMAGE \cite{EMAGE} used body cues to enhance expression. DiffSHEG \cite{31} introduced a diffusion-based framework with unidirectional flow from expression to gesture, improving coordination but limiting mutual influence. Combo \cite{combo} attempted bidirectional modeling but suffered from synchronous generation inefficiencies. Our work addresses these issues with a bidirectional yet asynchronous framework, improving both interaction quality and generation efficiency without compromising diversity.

\begin{figure}
\includegraphics[width=\textwidth]{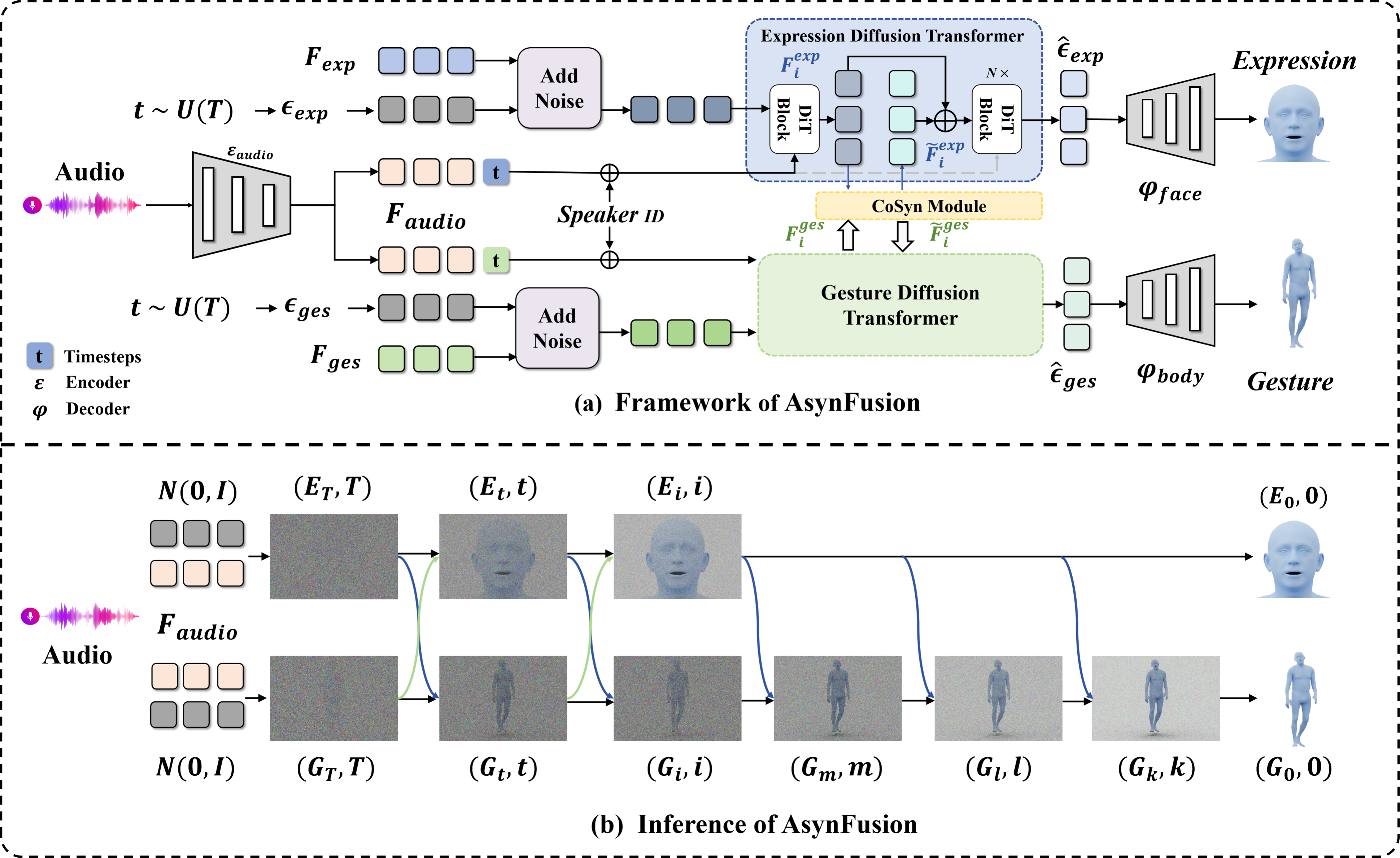}
\caption{Overview of AsynFusion. The framework (a) consists of a Dual-branch DiT architecture (blue and green) with a CoSync module for bidirectional feature interaction between expression and gesture branches, utilizing $F_i^{exp}$ and $F_i^{ges}$. (b) is the inference scheduler of AsynFusion.} \label{fig2}
\end{figure}

\section{Method}

\subsection{Preliminary}

Before introducing the details of our framework, we first present two key technical foundations that underpin our approach: \textbf{Diffusion Transformers (DiT)}  \cite{50} and \textbf{Latent Consistency Models (LCM)} \cite{lcm}. These preliminaries provide the necessary groundwork for understanding the design and implementation of our model.\\

\noindent\textbf{Diffusion Transformers}
Diffusion Transformers (DiT) employ a latent diffusion model (LDM) with a Transformer backbone for motion generation. Let $\mathbf{x}_0^{E}$, $\mathbf{x}_0^{G}$, and $\mathbf{x}_0^M$ denote expressions, gestures, and motion clips, respectively, aiming to model the motion distribution $p(\mathbf{x}_0) \in \mathbb{R}^{N \times (3J + D_{exp})}$, where $N$ is the number of frames, $J$ the skeletal joints, and $D_{exp}$ the expression blend shape dimension.

The input motion $\mathbf{x}_0 \sim p(\mathbf{x}_0)$ is progressively corrupted over $T$ steps via a noise schedule $\beta_t \in (0,1)$:
\begin{equation}
q(\mathbf{x}_t \mid \mathbf{x}_{t-1}) = \mathcal{N}\left(\mathbf{x}_t; \sqrt{1 - \beta_t}\mathbf{x}_{t-1}, \beta_t \mathbf{I}\right),
\end{equation}
with the full process given as:
\begin{equation}
q(\mathbf{x}_{1:T} \mid \mathbf{x}_0) = \prod_{t=1}^{T} q(\mathbf{x}_t \mid \mathbf{x}_{t-1}).
\end{equation}

Starting from Gaussian noise $\mathbf{x}_T$, the denoising process iteratively refines $\mathbf{x}_t$, using a transformer-based noise predictor $\epsilon_\theta$ to estimate the noise. Following DDPM \cite{16}, the denoised sample $\mathbf{x}_{t-1}$ is computed as:
\begin{equation}
\mathbf{x}_{t-1} = \frac{1}{\sqrt{\alpha_t}}\left(\mathbf{x}_t - \frac{\beta_t}{\sqrt{1-\bar{\alpha}_t}} \epsilon_\theta(\mathbf{x}_t,t)\right) + \sigma_t\mathbf{z}, 
\end{equation}

where $\mathbf{z} \sim \mathcal{N}(0,\mathbf{I})$, and $\alpha_t$, $\beta_t$, $\sigma_t$ are time-dependent coefficients.

For conditional generation, the noise predictor also takes a conditioning signal $c$ (e.g., audio features) as input, expressed as:
\begin{equation}
\epsilon_\theta(\mathbf{x}_t,t,c) = f_{i} \circ f_{i-1} \circ \dots \circ f_1(\mathbf{x}_t,t,c),
\end{equation}

where each $f_i$ represents a DiT block.\\

\noindent\textbf{Latent Consistency Models}

Latent Consistency Models (LCM) accelerate diffusion sampling by learning a direct mapping from noisy latents to denoised results in fewer steps. Given a motion distribution $p(\mathbf{x}_0)$, the forward process is defined as:

\begin{equation}
    q(\mathbf{x}_t \mid \mathbf{x}_0) = \mathcal{N} \left(\mathbf{x}_t; \sqrt{\bar{\alpha}_t} \mathbf{x}_0, (1 - \bar{\alpha}_t) \mathbf{I} \right),
\end{equation}

where $\bar{\alpha}_t = \prod_{i=1}^{t} \alpha_i$ and $\alpha_i = 1 - \beta_i$.
The key idea of LCM is to learn a consistency model $f_\theta$ that directly estimates the clean sample $\mathbf{x}_0$ from $\mathbf{x}_t$:

\begin{equation}
    f_\theta(\mathbf{x}_t, t) \approx \mathbb{E}_{q(\mathbf{x}_0 \mid \mathbf{x}_t)}[\mathbf{x}_0].
\end{equation}
This allows for faster sampling compared to traditional diffusion models. The training objective for LCM is:

\begin{equation}
    \mathcal{L}_{LCM} = \mathbb{E}_{\mathbf{x}_0, t, \epsilon} \left[ \left| \mathbf{x}_0 - f_\theta\left(\sqrt{\bar{\alpha}_t} \mathbf{x}_0 + \sqrt{1 - \bar{\alpha}_t} \epsilon, t\right) \right|^2 \right].
\end{equation}
For conditional generation, the consistency model is augmented with the conditioning signal $c$, leading to $f_\theta(\mathbf{x}_t, t, c)$. This forms the basis for the asynchronous sampling strategy in our framework.

\subsection{The AsynFusion Framework}
{\bf Overview.}  As shown in Fig. \ref{fig2} (a), we propose \textbf{AsynFusion}, a novel framework designed to synthesize coordinated facial expressions and body gestures in real-time. Traditional methods often rely on cascaded architectures or simple feature fusion strategies, which can compromise both computational efficiency and the fidelity of motion synthesis. Recent unified frameworks enforce unidirectional information flow, limiting the dynamic interplay between expressions and gestures, which is essential for natural human communication. To address these challenges, AsynFusion introduces a {Dual-branch DiT Architecture} to process expressions and gestures in parallel, a {Cooperative Synchronization (CoSync) Module} to dynamically synchronize the two branches at different sampling rates during inference, and an {Asynchronous LCM Sampling} framework to accelerate the sampling process, enabling real-time synthesis of coordinated motion. Fig. \ref{fig2} (b) shows the inference scheduler of AsynFusion with asynchronous LCM sampling.\\
Next, we will  detail the {Dual-branch DiT Architecture}, {CoSync Module}, and {Asynchronous LCM Sampling}.\\

\noindent{\bf Dual-branch DiT Architecture.}  
To generate coordinated facial expressions and body gestures, AsynFusion adopts a {\it Dual-branch DiT Architecture} consisting of two parallel branches: the \textit{expression branch}, which captures subtle facial motions and lip synchronization, and the \textit{gesture branch}, which handles broader body dynamics. Each branch uses independent transformer blocks to learn domain-specific temporal dependencies. The input to each branch includes: (1) noisy motion samples $\mathbf{z}_t^E$ or $\mathbf{z}_t^G$, obtained by adding Gaussian noise to target motions; (2) timestep embeddings $\gamma(t)$; and (3) shared audio features $\mathbf{F}_{\mathrm{aud}}$ for synchronized conditioning:
\begin{equation}
\hat{\mathbf{z}}^E  = \mathcal{T}_E(\mathbf{z}_t^E, \gamma(t), \mathbf{F}_{\mathrm{aud}}), \quad
\hat{\mathbf{z}}^G  = \mathcal{T}_G(\mathbf{z}_t^G, \gamma(t), \mathbf{F}_{\mathrm{aud}}),
\end{equation}
where $\mathcal{T}_E$ and $\mathcal{T}_G$ are the expression and gesture transformers. This design supports specialized learning per modality while enabling cross-branch interaction through synchronization. It also allows for asynchronous sampling to accommodate their differing temporal characteristics.\\

\noindent{\bf Cooperative Synchronization Module.}  
To model the interplay between facial expressions and gestures, we introduce the Cooperative Synchronization (CoSync) module, which enables bidirectional feature exchange between branches. After each transformer block, a cross-attention-based synchronization layer captures inter-modal dependencies and enhances motion coherence.\\
We take gesture to expression data-flow for example, the query $\mathbf{Q}_{exp}$ is extracted by linear projection from $\mathbf{F}_i^{exp}$ ($i$ is the layer index), and the key and value $\mathbf{K}_{ges}$, $\mathbf{V}_{ges}$ are extracted from $\mathbf{F}_i^{ges}$ in the same way. To obtain the updated facial feature $\tilde{\mathbf{F}}_i^{exp}$,

\begin{equation}
\mathbf{F}_i^{ges \to exp}=\text{softmax}\left( \frac{\mathbf{Q}_{exp} (\mathbf{K}_{ges})^\top}{\sqrt{d}} \right) \mathbf{V}_{ges}, \label{6}
\end{equation}

\begin{equation}
\tilde{\mathbf{F}}_i^{exp}=\text{MLP}(\text{LN}(\mathbf{F}_i^{ges \to exp})) + \mathbf{F}_i^{exp}, \label{7}
\end{equation}

where MLP and LN is a MLP block and a LayerNorm, $\sqrt{d}$ is a scaling factor.
This bidirectional feature exchange enables the model to capture subtle correlations between facial micro-expressions and corresponding gestural nuances, much like the natural synchronization observed in human behavior. What distinguishes the CoSync module is its ability to maintain the delicate balance between modality-specific independence and cross-modal coordination. While each branch preserves its specialized focus, the module enables them to share complementary information that enhances the overall coherence of the generated animation.\\

\noindent{\bf Asynchronous LCM Sampling.}
To achieve efficient real-time generation while preserving the benefits of bidirectional interaction, we introduce an asynchronous sampling strategy based on Latent Consistency Models (LCM). Specifically, we train separate LCM models for the expression and gesture branches, each optimized for their respective sampling step:
\begin{equation}
\begin{aligned}
f_{\theta_{exp}}(\mathbf{x}_t^E, t) \approx \mathbb{E}{q(\mathbf{x}_0^E|\mathbf{x}_t^E)}[\mathbf{x}_0^E]\\
f_{\theta_{ges}}(\mathbf{x}_t^G, t) \approx \mathbb{E}{q(\mathbf{x}_0^G|\mathbf{x}_t^G)}[\mathbf{x}_0^G]
\end{aligned}
\end{equation}
The expression branch typically requires fewer sampling steps ($T_{exp}$) than the gesture branch ($T_{ges}$) due to its more constrained motion space. To support bidirectional interaction during asynchronous sampling, we introduce a dynamic feature buffer in the CoSync module. At each step, both branches store and asynchronously update their intermediate features. This allows each branch to access the latest features from the other, maintaining continuous cross-modal exchange despite differing sampling rates. As a result, the expression branch achieves fast generation for facial motions, while the gesture branch uses more steps to capture complex dynamics—balancing quality and efficiency.

\subsection{Training}
Our training framework optimizes separate loss functions for both the expression and gesture branches, ensuring high-quality motion generation in each domain. While each branch is trained independently, interaction is maintained through the CoSync module. The loss components for both branches (expression $E$ and gesture $G$) are as follows.

First, the noise prediction loss $\mathcal{L}_t$ is defined as:
\begin{equation}
\mathcal{L}_t = \mathbb{E}_{\mathbf{x}_0, \epsilon} \left[ \|\epsilon - \epsilon_\theta(\sqrt{\bar{\alpha}_t}\mathbf{x}_0 + \sqrt{1 - \bar{\alpha}_t} \epsilon, t) \|^2 \right]
\label{noise_loss}
\end{equation}
This loss predicts the noise added during the diffusion process.

Next, the velocity loss $\mathcal{L}_v$ is computed to measure the difference in velocity between the ground-truth motion $\mathbf{x}_0$ and the predicted motion $\hat{\mathbf{x}}_0$. To compute the velocity difference, we first derive the predicted motion $\hat{\mathbf{x}}_0$ from the predicted noise $\hat{\epsilon}_t$. The velocity loss is then given by:
\begin{equation}
    \mathcal{L}_v = \mathbb{E} \left[ \| (\mathbf{x}_0[1:] - \mathbf{x}_0[:-1]) - (\hat{\mathbf{x}}_0[1:] - \hat{\mathbf{x}}_0[:-1]) \|^2 \right]
    \label{velocity_loss}
\end{equation}

Finally, we use the Huber loss $\mathcal{L}_\delta$ for motion reconstruction. This loss is defined as:
\begin{equation}
\mathcal{L}_\delta = \begin{cases}
\frac{1}{2}(\mathbf{x}_0 - \hat{\mathbf{x}}_0)^2, & \text{if } |\mathbf{x}_0 - \hat{\mathbf{x}}_0| < \delta, \\
\delta(|\mathbf{x}_0 - \hat{\mathbf{x}}_0| - \frac{1}{2} \delta), & \text{otherwise.}
\end{cases}
\end{equation}

The final loss is a weighted sum of the three losses:
\begin{equation}
\mathcal{L} = \lambda_t \mathcal{L}_t + \lambda_v \mathcal{L}_v + \lambda_\delta \mathcal{L}_\delta
\end{equation}

where the weights are set as $\lambda_t = 10$, $\lambda_v = 1$, and $\lambda_\delta = 1$ in our experiments.

\subsection{Long Sequence Generation}

To generate arbitrary-length animations, we further integrate existing techniques in DiffSHEG \cite{31} with our dual-branch asynchronous framework. The main challenge is ensuring smooth transitions between clips while maintaining the efficiency of asynchronous sampling.\\
\textbf{Clip-based generation:} We use a sliding window approach similar to DiffSHEG. Consecutive clips have overlapping frames to ensure smooth transitions. The starting frames of each new clip are initialized with the ending frames of the previous clip, maintaining continuity in both facial expressions and body gestures. This approach naturally fits with our asynchronous sampling mechanism, allowing each branch to sample at its own optimal rate while preserving temporal coherence.\\
\textbf{Efficient inference:} Our LCM-based method significantly reduces the number of required sampling steps compared to traditional diffusion models. The expression branch typically needs 4-6 sampling steps, while the gesture branch uses 6-8 steps—both far fewer than the 1000 steps often used in conventional models. This reduction in steps is crucial for real-time avatar animation applications. It is noteworthy that although our method employs asynchronous generation—meaning the time required to generate facial expressions and body poses for each frame may vary—the outputs are produced on a frame-by-frame basis, ensuring that the expressions and poses are aligned in each frame. \\
\textbf{Transition refinement:} To ensure smooth transitions between clips, we apply a transition refinement technique. Overlapping frames at the clip boundaries are linearly interpolated during the final sampling steps of each branch, ensuring seamless transitions while keeping expressions and gestures naturally coordinated.

\section{Experiments}
\subsection{Datasets}
We evaluate our method on three public speech-motion datasets. \textbf{BEAT Dataset} \cite{26} provides synchronized speech, facial expressions, and gestures from four subjects, along with annotations such as transcriptions, semantics, and emotions. Following the official setup, we use 34-frame clips for training/validation and 64-frame (approx. one minute) sequences for testing. Motions are represented using axis-angle rotation at 15 FPS. \textbf{SHOW Dataset} \cite{show} offers synchronized SMPLX \cite{33} parameters and audio (22 kHz) from four speakers, recorded at 30 FPS. We use 88-frame sequences for training/validation and variable-length clips for testing, with SMPLX as the motion representation.

\subsection{Metrics Computation}

This section outlines the evaluation metrics employed in our experimental analysis.\\
{\bf Fréchet Motion Distance} The Fréchet Motion Distance (FMD) \cite{sup1} extends the established Fréchet Gesture Distance concept, providing a reliable measure that aligns with human perceptual assessment. FMD quantifies the distributional similarity between generated and authentic motions by computing the Fréchet distance between their respective latent representations. These latent features are obtained through a specialized neural encoder trained on either the BEAT\cite{26} or SHOW datasets\cite{show}. The mathematical formulation is:

\begin{equation}
   FGD = |\mu_r - \mu_s|^2 + Tr(\sigma_r + \sigma_s - 2\sqrt{\sigma_r\sigma_s}),
\end{equation}

where $(\mu_s, \sigma_s)$ represent the mean and covariance statistics of the synthesized motion distribution in latent space, while $(\mu_r, \sigma_r)$ correspond to those of the real motion distribution. Following this framework, we define analogous metrics - Fréchet Expression Distance (FED) and Fréchet Gesture Distance (FGD) - to evaluate expression and gesture quality respectively.\\

{\bf Diversity metric (Div)} To assess the variability of generated animations, we employ a diversity metric \cite{5} that quantifies motion heterogeneity across batches. Given a test batch dimension B, we calculate our diversity score as:

\begin{equation}
   Div = \frac{2}{B \times (B-1)} \sum_{i=1}^{B-1} \sum_{j=i+1}^{B} |\hat{x}_i - \hat{x}_j|_1,
\end{equation}

For implementation, $\hat{x}_i$ represents a complete motion sequence from our i-th batch generation. In our experimental protocol, we utilize a batch size B of 50 samples to ensure robust diversity assessment.\\
{\bf Beat Alignment (BA)} To evaluate temporal coherence between audio and generated movements, we implement the Beat Alignment (BA) metric. This assessment tool examines the temporal correlation by quantifying the proximity between motion-derived beats and their audio counterparts. The mathematical representation is:

\begin{equation}
   BA = \frac{1}{n}\sum_{i=1}^n \exp{-\frac{\min_{\forall b_j^a\in B^a} |b_i^m - b_j^a|^2}{2\sigma^2}},
\end{equation}

In this formulation, $B^a$ represents the set of detected audio beats $\{b_i^m\}$, while $B^m$ encompasses the extracted motion beats $\{b_i^m\}$. Our implementation adheres to the standardized beat detection and alignment procedures established in the BEAT \cite{26} and TalkSHOW \cite{show}.

\subsection{Implementation Details}
Experiments are conducted on two NVIDIA A100 (40G) GPUs. On BEAT, we train for 1,000 epochs with a batch size of 1,600. On SHOW, due to longer sequences and higher frame rates, we train for 1,600 epochs with a batch size of 700. We compare AsynFusion with recent state-of-the-art methods, focusing on DiffSHEG \cite{31} and Combo \cite{combo} for joint generation. All models use axis-angle rotation and are conditioned on audio and speaker identity. For gesture-only models (DiffGesture \cite{DiffGesture}, DiffuseStyleGesture \cite{Diffusestylegesture}, LDA \cite{1}), we train separate facial models for fair comparison. Although our focus is on upper-body motion, AsynFusion supports full-body synthesis. We pay particular attention to comparing our bidirectional interaction and asynchronous sampling with the unidirectional design of DiffSHEG \cite{31}.

\begin{figure}
\vspace{-0.5cm}
  \centering
    \includegraphics[width=\textwidth]{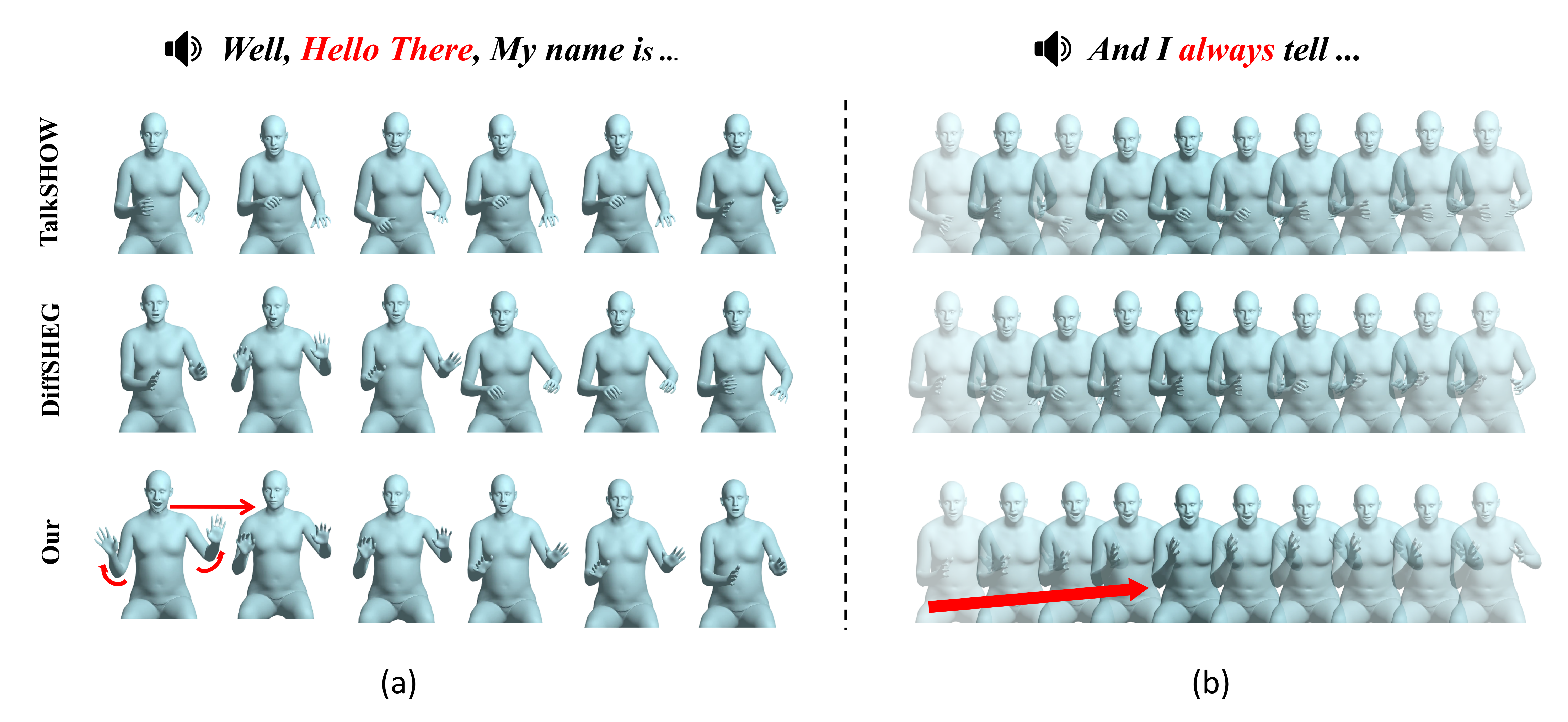}
    \caption{Visualization of generated motions for the speech. The red arrows indicate how the gestures and facial expressions are well-coordinated during the greeting motion.}
     \label{fig:vis}
\vspace{-0.5cm}
\end{figure}
\subsection{Qualitative Evaluation}

Fig. ~\ref{fig:vis} presents visualizations of motions generated by our method. In (a), for the speech {\bf ``Well, Hello There, My name is..."}, our model produces synchronized hand-raising gestures and facial expressions, demonstrating the effectiveness of bidirectional interaction. Notably, during “Hello There”, the greeting gesture aligns with a smooth transition from neutral to friendly expressions, and gesture peaks match speech emphasis. In (b), for \textbf{``And I always tell..."}, AsynFusion outperforms baselines by generating a rising hand gesture synchronized with the emphasis on “always”, as indicated by the red intensity curve. The facial expression shifts accordingly, showcasing cohesive non-verbal emphasis enabled by our bidirectional design.

\begin{table}
\caption{Quantitative comparison and ablation study on BEAT \cite{24}, SHOW \cite{show} datasets. Best results in each
category are in \textbf{bold}; second best are
\underline{underlined}.}
\label{tab:comparison}
\centering
\vspace{-0.1in}
\begin{tabular}{cl|c|cc|ccc}
\toprule
\multirow{2}{*}{Dataset}& \multirow{2}{*}{Method} & Holistic & \multicolumn{2}{c|}{Expression} & \multicolumn{3}{c}{Gesture} \\
\cmidrule{3-8}
& & FMD $\downarrow$ & FED $\downarrow$ & Div $\uparrow$ & FGD $\downarrow$ & BA$\uparrow$ & Div $\uparrow$ \\
\midrule
\multirow{7}{*}{BEAT \cite{26}} & Ground Truth & - & - & 0.651 & - & 0.915 & 0.819 \\
\cmidrule{2-8}
& CaMN \cite{26} & 1055.52 & 1324.00 & 0.479 & 1635.44 & 0.793 & 0.633 \\
& DiffGesture \cite{DiffGesture} & 12142.70 & 586.45 & 0.625 & 23700.91 & \underline{0.929} & \underline{3.284}\\
& DSG \cite{Diffusestylegesture} & 1261.59 & 998.25 & \underline{0.688} & 1907.58 & 0.919 & \textbf{0.701}\\
& LDA \cite{1} & 688.25 & 510.345 & 0.603 & 997.62 & \textbf{0.923} & 0.688\\
& DiffSHEG\cite{31} & \underline{324.67} & \underline{331.72} & 0.539 & \underline{438.93} & 0.914 & 0.536\\
\cmidrule{2-8}
& ours & \textbf{312.46} & \textbf{316.97} & 0.565 & \textbf{421.58} & 0.917 & 0.561\\
\midrule
\multirow{7}{*}{SHOW \cite{19}} & CaMN \cite{26} & 3.365 & - & - & 2.199 & 0.7998 & 10.13\\
& DSG \cite{Diffusestylegesture} & 3.462 & - & - & 2.404 & 0.8295 & 10.04\\
& TalkSHOW \cite{show} & 3.478 & - & - & 2.462 & 0.8449 & 10.29\\
& ProbTalk \cite{probtalk} & 3.980 & 5.59 & - & 5.21 & 0.8531 & 10.45\\
& EMAGE \cite{EMAGE} & 3.380 & - & - & 2.255 & 0.8585 & \underline{12.40}\\
& Combo \cite{combo} & \underline{3.142} & - & - & \underline{2.067} & \underline{0.8667} & 10.36\\
\cmidrule{2-8}
& ours & \textbf{3.098} & - & - & \textbf{2.049} & \textbf{0.8701} &\textbf{12.53}\\
\bottomrule
\end{tabular}
\vspace{-0.5cm}
\end{table}

\subsection{Quantitative Evaluation}

We evaluate our method using FMD, FGD, FED, diversity (Div), and beat alignment (BA) to capture motion quality, expressiveness, and temporal alignment. As shown in Table~\ref{tab:comparison}, AsynFusion consistently outperforms prior methods on both BEAT and SHOW datasets. On BEAT, it achieves the best FMD (312.46), FED (316.97), and FGD (421.58), indicating superior overall quality and gesture stability. It also improves gesture diversity (Div = 0.561) and beat alignment (BA = 0.917), closely matching ground truth. On SHOW, AsynFusion achieves top scores in FMD (3.098), Div (12.53), BA (0.8701), and FGD (2.049), surpassing Combo, TalkSHOW, and Probotalk across all metrics.\\

\noindent{Overall, AsynFusion sets a new benchmark for coordinated expression-gesture generation, offering superior motion stability, diversity, and synchronization through bidirectional feature interaction and asynchronous sampling.}

\section{Ablation Study}

\begin{table}
\caption{Ablation study on different architectural variants of our bidirectional feature interaction mechanism. The results demonstrate the effectiveness of our full model with bidirectional design.Best results in each
category are in \textbf{bold}; second best are \underline{underlined}.}\label{tab:ablation_arch}
\centering
\begin{tabular}{c|c|c|c}
\toprule
\multirow{2}{*}{Model} & \multicolumn{1}{c|}{Holistic} & \multicolumn{1}{c|}{Expression} & \multicolumn{1}{c}{Gesture} \\
\cmidrule{2-4}
& FMD $\downarrow$ & FED $\downarrow$ & FGD $\downarrow$ \\
\midrule
No interaction  & 352.14 & 341.56 & 471.42  \\
Uni-Flow ($E \rightarrow G$)  & \underline{321.37} & \underline{327.94} & \underline{435.24}  \\
Uni-Flow ($G \rightarrow E$)  & 343.72 & 342.58 & 457.83 \\
Naïve Fusion  & 340.16 & 340.23 & 467.33 \\
CoSync ($G \leftrightarrow E$) & \textbf{312.46} & \textbf{316.97} & \textbf{421.58} \\
\bottomrule
\end{tabular}
\end{table}
In this section, we conduct comprehensive ablation studies to validate the effectiveness of AsynFusion's key components and design choices on BEAT Dataset \cite{26}. Specifically, we examine (1) the impact of different feature interaction strategies in our Dual-branch DiT Architecture, (2) the efficiency of our asynchronous sampling approach compared to synchronized alternatives.\\

 \noindent\textbf{Impact of Feature Interaction Strategies.} To evaluate interaction mechanisms between expression and gesture branches, we compare five variants: (1) No Interaction, (2) Unidirectional Flow ($E \rightarrow G$), (3) Unidirectional Flow ($G \rightarrow E$), (4) Naïve Fusion, and (5) our Bidirectional Interaction. As shown in Table~\ref{tab:ablation_arch}, the Bidirectional Interaction significantly outperforms all others (FMD = 312.46, FED = 316.97, FGD = 421.58). The No Interaction baseline yields poor coordination (FMD = 352.41), highlighting the need for cross-branch communication. Unidirectional flows improve performance but exhibit modality imbalance—$E \rightarrow G$ favors facial metrics (FGD = 435.24), while $G \rightarrow E$ performs worse than Naïve Fusion. Naïve Fusion enables information sharing but fails to capture modality dynamics (FMD = 340.16). These results support our bidirectional design and motivate the proposed asynchronous LCM sampling strategy.\\
 
\noindent\textbf{Impact of Different Feature Fusion Methods.} In our exploration of improving the coordination between facial expressions and body gestures, we explored three distinct feature fusion approaches: {\bf Cross Attention (CA)}, {\bf Feature Concatenation (FC)}, {\bf Gated Fusion (GF)}. Our AsynFusion framework implements Cross Attention, leveraging its bidirectional mechanism to achieve fine-grained control over expression-gesture information flow, thereby producing more natural and expressive animations. We also examined two alternative methods: Feature Concatenation, which simply concatenates expression and gesture features before feeding them into the next DiT block, and Gated Fusion, which employs learnable weights through a gating mechanism to control the fusion process:
\begin{equation}
\begin{aligned}
F_{fused} = \sigma(W_g[F_{exp};F_{ges}]) \odot F_{exp} + \\
(1-\sigma(W_g[F_{exp};F_{ges}])) \odot F_{ges}
\end{aligned}
\end{equation}
where $\sigma$ represents the sigmoid function, and $W_g$ are the learnable weights. As shown in Table~\ref{tab:ablation_fusion} This approach adaptively controls each modality's influence through learnable weights. While Gated Fusion achieves better modality balance than simple concatenation, it lacks the sophisticated bidirectional interaction of Cross Attention, making it less capable of capturing the subtle dependencies in natural human behavior.

\begin{table}[t]\small
\caption{Ablation study on different Fusion Strategy. Best results in each
category are in \textbf{bold}; second best are \underline{underlined}.}\label{tab:ablation_fusion}
\centering
\begin{tabular}{c|c|c|c}
\toprule
\multirow{2}{*}{Fusion Strategy} & \multicolumn{1}{c|}{Holistic} & \multicolumn{1}{c|}{Expression} & \multicolumn{1}{c}{Gesture} \\
\cmidrule{2-4}
& FMD $\downarrow$ & FED $\downarrow$ & FGD $\downarrow$ \\
\midrule
Feature Concat & 340.16 & 340.23 & 467.33 \\
Gated Fusion & \underline{325.46} & \underline{329.15} & \underline{443.74}  \\
Cross Attention & \textbf{312.46} & \textbf{316.97} & \textbf{421.58} \\
\bottomrule
\end{tabular}
\end{table}

 \noindent\textbf{Efficiency of Asynchronous LCM Sampling.} We evaluate three sampling strategies: (1) DDIM (25 steps), (2) Synchronized LCM (8 steps), and (3) our Asynchronous LCM (4 steps for expression, 8 for gesture). As shown in Table~\ref{tab:lcm}, DDIM achieves the best quality (FMD = 312.46) but is slow (56.4s). Synchronized LCM reduces time by 67\% (18.6s) with minimal quality drop (FMD = 318.13). Our Asynchronous LCM further improves efficiency (15.9s, 72\% faster than DDIM) while maintaining competitive quality (FMD = 320.59), showing the benefit of adapting sampling to the convergence speed of each branch.\\

\begin{table}[t]\small
\caption{Comparison of different sampling strategies.}\label{tab:lcm}
\centering
\begin{tabular}{l|c|c|c}
\toprule
Sampling Strategy & Steps (E/G) & Time (s) & FMD $\downarrow$ \\
\midrule
w/o LCM & 25/25 & 56.4 & \textbf{312.46} \\
Sync LCM & \underline{8/8} & \underline{18.6} & \underline{318.13} \\
Async LCM & \textbf{4/8} & \textbf{15.9} & 320.59 \\
\bottomrule
\end{tabular}
\end{table}

\section{Conclusion}
A key limitation of AsynFusion lies in its dependency on training data quality. Our model, like other deep learning approaches, inherits both desired behaviors and undesirable artifacts from the training datasets. For example, when trained on BEAT \cite{26}, the generated motions exhibit jittering artifacts similar to those in the dataset's female character animations. Similarly, expression discontinuities from the SHOW dataset appear in our synthesized results. These observations highlight that future improvements in motion synthesis may rely as much on better data collection and cleaning as on model architecture advances. Future research could focus on end-to-end multimodal foundation models for motion synthesis, processing speech, text, and video simultaneously. Large-scale pretraining across diverse data sources could enable deeper understanding of verbal and non-verbal communication patterns. This approach could enhance motion diversity and naturality through universal representations, while advanced cross-modal pretraining could improve the capture of speech-motion correlations for more nuanced animations.

\section*{Acknowledgement}
\textbf{This work was supported by the National Natural Science Foundation of China under Grant No. 62441617 and 62476224. It was supported by the Postdoctoral Fellowship Program and China Postdoctoral Science Foundation under Grant No. 2024M764093 and Grant No. BX20250485, the Beijing Natural Science Foundation under Grant No. 4254100, the Fundamental Research Funds for the Central Universities under Grant No. KG16336301, and by Beijing Advanced Innovation Center for Future Blockchain and Privacy Computing.}
\end{document}